\documentclass[twocolumn,showpacs,prl,amsmath,amssymb,floatfix]{revtex4}
\usepackage{amsmath,amssymb}
\usepackage{bm}
\usepackage{epsfig}
\usepackage{psfrag}

\newcommand{\bd}{\bm}

\begin{document}

\title{Confined coherence in quasi-one-dimensional metals}

\author{Sascha Ledowski and Peter Kopietz}
  
\affiliation{Institut f\"{u}r Theoretische Physik, Universit\"{a}t
  Frankfurt,  Max-von-Laue Strasse 1, 60438 Frankfurt, Germany}

\date{March 28, 2007}

 \begin{abstract}

We present a functional renormalization group calculation
of the effect of strong interactions on the
shape of the Fermi surface  of weakly coupled  metallic chains.
In the regime where the bare interchain hopping is small, 
we show that scattering processes involving large momentum
transfers perpendicular to the chains
can completely destroy  the warping of the true Fermi surface, 
leading to a confined state where
the renormalized interchain hopping vanishes and a
coherent motion perpendicular to the chains is impossible.

\end{abstract}


\maketitle

\begin{figure}[tb]
  \centering
  \epsfig{file=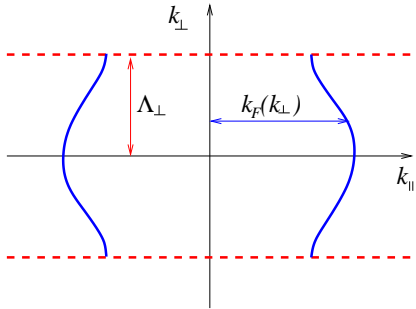,width=55mm}
  \vspace{-4mm}
  \caption{%
(Color online) 
FS of a two-dimensional array of weakly coupled metallic chains.
The dashed lines mark the boundary of the first
Brillouin zone in the direction perpendicular to the chains.
}
  \label{fig:FS}
\end{figure}

{\it{Introduction.}} 
 Electron-electron interactions can strongly modify the
Fermi surface (FS) of a metal.  A well known example is the
Pomeranchuk transition, where
the symmetry of the FS is spontaneously broken due
to strong interactions in the zero-sound channel~\cite{Pomeranchuk58}.
However, there are other quantum phase transitions 
associated with the geometry or the topology
of the FS without symmetry breaking, such as the
Lifshitz transition \cite{Lifshitz60,Quintanilla06} 
or the truncation transition \cite{Furukawa98,Ferraz03}, where certain sectors of the FS
are washed out by interactions, while others remain intact.
Another example is the interaction-induced confinement transition, 
which has been  proposed by Clarke, Strong, and Anderson more than 
ten years ago~\cite{Clarke94}:  they considered
metallic chains with small interchain hopping $t_{\bot , 0}$.
For weak interactions, the FS 
consists then of two disconnected weakly curved sheets as shown
Fig.~\ref{fig:FS}.
%
%
%
%
The amplitude of the warping of the FS
is proportional to the renormalized interchain hopping $t_{\bot}$.
Clarke {\it{et al.}} \cite{Clarke94} suggested that, at least for 
sufficiently strong interactions and small $t_{\bot,0}$, 
the renormalized $t_{\bot}$ vanishes. 
The true FS is then completely flat, so that a coherent motion
of the electrons in the direction perpendicular to the chains 
is not possible (confined coherence).

In the past decade the confinement problem in weakly coupled metallic chains
has been studied by many 
authors \cite{Clarke94,Wen90,Kopietz94,Boies95,Arrigoni98}, 
but the results have not converged due to a lack of controlled methods.
A simple one-loop calculation \cite{Wen90,Boies95} suggests that
the renormalized interchain hopping vanishes  if the anomalous dimension
$\eta$ of the Luttinger liquid state without interchain hopping is
larger than unity. However, this argument does not take the renormalization of $\eta$
by interchain hopping into account.
Indeed, more refined calculations by Arrigoni \cite{Arrigoni98}
suggest that higher order corrections are important and possibly lead to a finite
$t_{\bot}$ even for $\eta > 1$.
In this Letter we shall re-examine this problem using a novel
functional renormalization group (RG) approach involving
both fermionic and bosonic fields~\cite{Schuetz05,Ledowski07}.
Our main result is that 
the regime of confined coherence proposed in Ref.~\cite{Clarke94} indeed exists, so that
strong interactions can give rise to a non-Fermi liquid
normal state in quasi-one-dimensional metals.

{\it{Model.}} 
We start from an effective low energy model 
for spinless fermions with linearized energy dispersion and
density-density interactions. The  Euclidean action is
 \begin{eqnarray}
 S [ \bar{\psi} , \psi ] & = & \sum_{\alpha } \int_K \bigl[ - i \omega + \alpha v_F   \delta k_{\parallel}  + \mu_0 ( k_{\bot} ) \bigr]
 \bar{\psi}_{  K \alpha }  {\psi}_{  K \alpha } 
 \nonumber
\\
 & + & \frac{1}{2} \sum_{\alpha \alpha^{\prime}} \int_{\bar{K}} 
f_{  \alpha \alpha^{\prime} }
 \bar{\rho}_{ \bar{K} \alpha} \rho_{ \bar{K} \alpha^{\prime}}
 \; ,
 \label{eq:Sdef}
 \end{eqnarray} 
where
$ \mu_0 ( k_{\bot} ) = - \Sigma ( {\bd{k}}_F , i 0 )$ 
is a counter-term involving  the exact
self-energy at the true FS ${\bd{k}} = {\bd{k}}_F$ and
zero frequency, 
and $\delta k_{\parallel} = k_{\parallel} - \alpha k_F ( k_{\bot} )$.
Here $k_{\parallel}$ is the component of the two-dimensional
lattice momentum $\bd{k}$ parallel to the chain direction,
and
$k_{\bot}$ is the corresponding  perpendicular component.
The FS $\bd{k}_F$
can then be parameterized  as
$ k_{\parallel} = \alpha k_F ( k_{\bot})$, where
$\alpha = \pm 1$ labels the two disconnected sheets of the FS,
see Fig.~\ref{fig:FS}.
We neglect the $k_{\bot}$-dependence of the Fermi velocity $v_F$.
The chiral fields $ \psi_{ K  \alpha}$ are defined in terms of
the usual Fermi fields $\psi_{ k_{\parallel} , k_{\bot} , i \omega }$
via
$ \psi_{ K  \alpha} = \psi_{  \alpha k_{ F} ( k_{\bot} ) + \delta k_{\parallel} ,  
k_{\bot} , i \omega }$, and the
chiral densities $\rho_{ \bar{K}  \alpha}$
are $
\rho_{ \bar{K}  \alpha} = \int_K 
 \bar{\psi}_{K \alpha} \psi_{ K + \bar{K} \alpha}
 $.
We use collective labels
$K = ( \delta k_{\parallel} , k_{\bot} , i \omega )$ for fermionic    
and $\bar{K} = (  \bar{k}_{\parallel} , \bar{k}_{\bot} , i \bar{\omega} )$ for
bosonic fields, 
where $\omega$ and $\bar{\omega}$ are Matsubara frequencies.
The integration symbols are
 $
 \int_K =  \int_{k_{\bot}}
 \int_{ - \Lambda_\parallel}^{\Lambda_\parallel}  
 \frac{d \delta k_{\parallel}}{2 \pi}  \int  \frac{d \omega}{2 \pi}
$
and
 $
 \int_{\bar{K}} =   \int_{\bar{k}_{\bot}}
 \int_{ - \bar{\Lambda}_\parallel}^{\bar{\Lambda}_\parallel}  
\frac{d  \bar{k}_{\parallel}}{2 \pi}  \int  \frac{d \bar{\omega}}{2 \pi }
$ where for later convenience we have introduced the notation
$
\int_{ k_{\bot} }  =   \int_{ - \Lambda_{\bot} }^{  \Lambda_{\bot}}
 \frac{ d k_{\bot}}{ 2  \Lambda_\bot }$ and
$
\int_{ \bar{k}_{\bot} }  =  \int_{ - \bar{\Lambda}_{\bot}}^{ \bar{\Lambda}_{\bot} }
 \frac{ d \bar{k}_{\bot}}{  2 \bar{\Lambda}_{\bot} }
$.
Here $\Lambda_{\parallel}$ is a bandwidth cutoff,
$\Lambda_{\bot} = \pi / a_{\bot}$ is the width of the Brillouin zone
in transverse direction, and
$\bar{\Lambda}_{\bot}$ and $\bar{\Lambda}_\parallel$ 
restrict the momentum transfered by the interaction
in the directions parallel and perpendicular to the chains.
We assume that $ \bar{\Lambda}_{\parallel} \ll {\rm min} \{ k_F ( k_{\bot} )\}$,
so that the  interaction 
$f_{ \alpha \alpha^{\prime} }$ in Eq.~(\ref{eq:Sdef}) 
does not transfer momentum between the two disconnected sheets of the FS. 
However, the transverse 
momentum transfer cutoff $\bar{\Lambda}_{\bot}$ can be of the order of the 
transverse width  $\Lambda_{\bot} = \pi  / a_{\bot}$
of the Brillouin zone, so that
transverse Umklapp scattering is possible.
For simplicity we set $\bar{\Lambda}_{\bot} = \Lambda_{\bot}$ and call
$ \bar{\Lambda}_{\parallel} = \Lambda_0 $.

{\it{Self-consistent perturbation theory.}} 
To begin with, let us calculate the FS within second order 
self-consistent perturbation theory.  Using the procedure
outlined in Ref.~[\onlinecite{Neumayr03}], we obtain 
the following  integral equation for the
difference $\delta {k}_F ( k_{\bot} ) = k_{F} ( k_{\bot} ) - 
k_{F,0 } ( k_\bot )$ between the true Fermi momentum $k_F ( k_{\bot} )$ 
and the corresponding $k_{F,0} ( k_{\bot} )$ without interactions at the same density, 
 \begin{eqnarray}
 \frac{\delta {k}_F ( k_{\bot} )}{ \Lambda_0 } & = & 
    \left[ - {g}_4 + \frac{ {g}_4^2 + {g}_2^2}{2} \right] 
 \int_{ \bar{k}_{\bot} }  \tilde{\Delta} ( k_{\bot} , \bar{k}_{\bot} ) 
\nonumber
 \\
  &  & \hspace{-10mm} -  {g}_2^2 
 \int_{ \bar{k}_{\bot} }  \int_{ k_{\bot}^{\prime} }
J ( \tilde{\Delta} ( k_{\bot} , \bar{k}_{\bot} );  
\tilde{\Delta} ( k_{\bot}^{\prime} , \bar{k}_{\bot} ) )
\; ,
 \label{eq:IE}
 \end{eqnarray}
where
 $ \tilde{\Delta} ( k_{\bot} , \bar{k}_{\bot} ) = 
  [ k_F ( k_{\bot} ) - k_F ( k_{\bot} + \bar{k}_{\bot} )]/ {\Lambda}_{0}$, and
 \begin{equation}
 J ( \tilde{\Delta} ; \tilde{\Delta}^{\prime} ) =  
 \frac{ \tilde{\Delta} + \tilde{\Delta}^{\prime}}{4} \ln \left[
 \frac{ 4  - ( \tilde{\Delta} - \tilde{\Delta}^{\prime} )^2 }{ 
 ( \tilde{\Delta} + \tilde{\Delta}^{\prime} )^2} \right]
 \; .
 \label{eq:Ydef}
\end{equation}
The dimensionless couplings $g_2$ and $g_4$ are defined via
$2 \pi g_4 = \nu_0 f_{++} = \nu_0 f_{--}$ and
$2 \pi g_2 = \nu_0 f_{+-} = \nu_0 f_{-+}$, where 
the factor $\nu_0 = \Lambda_{\bot} ( \pi v_F )^{-1} = ( a_{\bot} v_F )^{-1}$ is introduced
for convenience.
We have solved Eq.~(\ref{eq:IE}) numerically, but for small $t_{\bot}$
we can also obtain an approximate analytic solution
using the fact that in this case
the dominant renormalization of the FS is due to
the logarithmic term in Eq.~(\ref{eq:Ydef}). 
Suppose that the bare FS is of the form
 $
 k_{F,0} ( k_{\bot} ) = \bar{k}_F + t_0 \cos ( k_{\bot} a_{\bot} )
 $
where $t_0 = 2 t_{\bot,0} / v_F \ll \Lambda_0$ 
and the average $\bar{k}_F$ is fixed by the total density. 
The renormalized FS is then given by
 $
 k_F ( k_{\bot} ) = \bar{k}_F + t  \cos ( k_{\bot} a_{\bot} ) + \ldots
 $, where $t$ is proportional to the renormalized nearest neighbor
interchain hopping,
and the ellipsis denotes higher harmonics corresponding to longer range hoppings.
From the  numerical solution of the integral equation
(\ref{eq:IE}) we find that  for $g_2 , g_4 \ll 1$ the higher harmonics
are indeed small. Then Eq.~(\ref{eq:IE}) can be reduced
to a transcendental equation for $t$, which to 
leading logarithmic order in $t/ \Lambda_0$ 
can be written as
$t / t_0 = [1 + R ( t  ) ]^{-1}$,
with
 \begin{equation}
 R ( t )  \approx   \frac{g_4}{2} -   \frac{{g}_4^2}{4}
    +  \frac{{g}_2^2}{2} 
 \ln (  {\Lambda}_0   / | t|  )   
\; .
 \label{eq:Rdef}
\end{equation}  
A similar relation has been obtained
previously \cite{Fabrizio93,Ledowski07} for the
difference between the Fermi momenta associated with the bonding and the
anti-bonding band in two coupled spinless chains. 
Note that to first order in the bare interaction
$ R ( t ) \propto {g}_4$, so that a repulsive interaction ${g}_4 >0$
reduces
the warping of the FS, while for ${g}_4 < 0$ the warping
of the FS is enhanced. However, for  $t_{\bot} \rightarrow 0$ the
logarithmic term proportional to  ${g}_2^2$ always dominates and
predicts an interaction-induced  reduction of the FS warping,
irrespective of  the sign of the interaction.

\begin{figure}[tb]
  \centering
  \epsfig{file=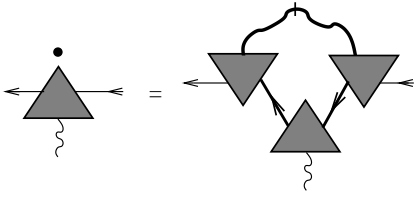,width=55mm}
  \vspace{-4mm}
  \caption{%
Diagrammatic representation of the flow equation (\ref{eq:gammaflow})
for the three-legged vertex with one bosonic (wavy line) and
two fermionic (solid lines with arrows) external legs.
The thick wavy line with a slash denotes the bosonic single scale propagator.
Additional contributions involving irrelevant higher order vertices
are omitted, see Refs.~[\onlinecite{Schuetz05,Ledowski07}].
}
\label{fig:threevertex}
\end{figure}

{\it{Functional RG approach.}}
We now generalize the RG approach developed in Ref.~[\onlinecite{Ledowski07}]
in the context of a simplified two-chain model  
to the more interesting two-dimensional case considered here.
The method has been  described in detail previously~\cite{Ledowski07}, 
so that we will be rather brief here.
In the momentum transfer cutoff scheme~\cite{Schuetz05} 
we decouple the density-density interaction by means of a
bosonic Hubbard-Stratonovich  transformation
and then use the maximal momentum carried by the  boson field
as flow parameter $\Lambda$ of the RG.
Our initial cutoff is thus
$\Lambda = \Lambda_0 = \bar{\Lambda}_{\parallel}$.
Eliminating boson fields with momenta in the range
$\Lambda < | \bar{k}_{\parallel} | < \Lambda_0$ 
we obtain a  new effective action, whose vertices
are determined by
a formally exact hierarchy of  functional RG flow equations.
To calculate the true FS, we need the flow equation
for the relevant part $r_l ( k_{\bot} )$
of the irreducible fermionic self-energy $\Sigma_{\Lambda} ( K , \alpha )$,
which is defined via
$r_l ( k_{\bot} ) = Z_l ( k_{\bot} )  [
\Sigma_{\Lambda} ( {\bd{k}}_F , i0, \alpha ) 
+ \mu_0 ( k_{\bot} ) ]/ v_F \Lambda $.  
Here $l = \ln ( \Lambda_0 / \Lambda )$ and
 $Z_l ( k_{\bot})$ is the flowing wave-function
renormalization. The functional RG flow equation for
$r_l ( k_{\bot} )$ is of the form
\begin{equation}
 \partial_l r_l ( k_{\bot} ) = [ 1 - \eta_l ( k_{\bot} ) ] r_l ( k_{\bot} )
 + \dot{\Gamma}_l ( k_{\bot} )
 \; ,
 \label{eq:flowr}
 \end{equation}
where $\eta_l ( k_{\bot} ) = - \partial_l \ln Z_l ( k_{\bot} )$
is the flowing anomalous dimension. 
An approximate expression
for the inhomogeneity $\dot{\Gamma}_l ( k_{\bot} )$ 
is given in Eq.~(\ref{eq:dotGamma}) below.
As long as $\eta_{\infty} ( k_{\bot} ) < 1$, the FS is well defined.
The shift $\delta k_F ( k_{\bot} )$ of the FS due to interactions can then be
obtained from the requirement that the initial value $r_0 ( k_{\bot} )$
should be fine tuned so that the relevant coupling
$r_{l} ( k_{\bot} )$ flows into a fixed point \cite{Kopietz01}.
This leads to the following exact integral equation for 
the FS,
 \begin{equation}
  \frac{ \delta k_F ( k_{\bot} ) }{\Lambda_0}
 = r_0 ( k_{\bot} ) =
 - \int_0^{\infty} d l e^{ -l + \int_0^l d t \eta_t ( k_{\bot} ) }
 \dot{\Gamma}_l ( k_{\bot} )
 \; .
 \label{eq:r0flow}
 \end{equation} 
Using the same truncation strategy as
in Ref.~[\onlinecite{Ledowski07}],
we approximate
 \begin{eqnarray}
  \dot{\Gamma}_l ( k_{\bot} )
 & = &
-   \int_{ \bar{k}_{\bot}} \int \frac{ d \bar{q} d \bar{\epsilon}}{(2 \pi )^2 }
\frac{  \delta ( | \bar{q} | -1 )   [ {\bf{F}}_l ( \bar{q} , i \bar{\epsilon} , 
 \bar{k}_{\bot}) ]_{\alpha \alpha}  e^{ i \bar{\epsilon} 0} 
 }{ i  \bar{\epsilon} 
 - \alpha  \bar{q} - \tilde{\Delta}_l ( k_{\bot} , \bar{k}_{\bot}) }
 \nonumber
\\
 &   & \times  
\gamma_l (  k_{\bot} , \bar{k}_{\bot} )
\gamma_l ( k_{\bot} + \bar{k}_{\bot}, - \bar{k}_{\bot} )
 \; ,
 \label{eq:dotGamma}
 \end{eqnarray}
where
 $
 \tilde{\Delta}_l ( k_{\bot} , \bar{k}_{\bot})  =  
\tilde{k}_{F,l} ( k_{\bot} ) - \tilde{k}_{F,l} ( k_{\bot} + \bar{k}_{\bot} )$
with $\tilde{k}_{F,l} ( k_{\bot} ) = k_F ( k_{\bot} ) / \Lambda - r_l ( k_{\bot} )$.
The inverse of the $2 \times 2$-matrix 
${\bf{F}}_l ( \bar{q} , i \bar{\epsilon} , \bar{k}_{\bot} )$
is defined via
 \begin{equation}
 [ {\bf{F}}_l ( \bar{q} , i \bar{\epsilon},
\bar{k}_{\bot} )]^{-1}_{\alpha \alpha^{\prime} }
 = [ \nu_0 {\bf{f}} ]^{-1}_{ \alpha \alpha^{\prime}} +
 \delta_{ \alpha \alpha^{\prime}} \tilde{\Pi}_l ( \bar{q} , i \bar{\epsilon},
\bar{k}_{\bot} , \alpha)
 \; ,
 \end{equation}
where $\bf{f}$ is a matrix in chirality space with
elements 
$[ {\bf{f}} ]_{\alpha \alpha^{\prime} } = f_{\alpha \alpha^{\prime} }$, and
$\tilde{\Pi}_l ( \bar{q} , i
 \bar{\epsilon} , \bar{k}_{\bot} , \alpha)$ is the rescaled polarization associated with
fermions of chirality $\alpha$, for which we use the
adiabatic approximation~\cite{Ledowski07}
 \begin{eqnarray}
\tilde{\Pi}_l ( \bar{q} , i \bar{\epsilon}, \bar{k}_{\bot}, \alpha ) & = & \frac{1}{2 \pi } 
 \int_{ k_{\bot}} 
 \frac{ \tilde{\Delta}_l ( k_{\bot} , \bar{k}_{\bot} ) + \alpha \bar{q} }{ \tilde{\Delta}_l ( k_{\bot} , 
 \bar{k}_{\bot} )   + \alpha \bar{q} - i \bar{\epsilon}    }
\nonumber
 \\\
 &   \times & \gamma_l (
 k_{\bot}  , \bar{k}_{\bot} )   \gamma_l (  
 k_{\bot} + \bar{k}_{\bot}  , - \bar{k}_{\bot} )
 \; .
 \label{eq:poladiabat}
\end{eqnarray}
The  anomalous dimension is in this approximation
 \begin{eqnarray}
 \eta_l ( k_{\bot} ) & = &
-   \int_{ \bar{k}_{\bot}} \int \frac{ d \bar{q} d \bar{\epsilon}}{(2 \pi )^2 }
\frac{  \delta ( | \bar{q} | -1 )  [ {\bf{F}}_l ( \bar{q} , i \bar{\epsilon} , 
 \bar{k}_{\bot}) ]_{\alpha \alpha} }{ [ i  \bar{\epsilon}
 - \alpha  \bar{q}  - \tilde{\Delta}_l ( k_{\bot} , \bar{k}_{\bot})]^2 }
 \nonumber
 \\
 &  &  \times 
\gamma_l (  k_{\bot} , \bar{k}_{\bot} )
\gamma_l ( k_{\bot} + \bar{k}_{\bot}, - \bar{k}_{\bot} )
 \; .
 \label{eq:etaflow}
 \end{eqnarray}
Finally, the dimensionless vertex 
$\gamma_l (  k_{\bot} , \bar{k}_{\bot} )$ 
with one bosonic and two fermionic external legs
(where $k_{\bot}$ labels the incoming fermion and 
$\bar{k}_{\bot}$ labels the boson) satisfies
the flow equation 
 \begin{eqnarray}
 \partial_l  \gamma_l ( k_{\bot} , \bar{k}_{\bot} )
 & = & - \frac{1}{2} [  \eta_l ( k_{\bot} ) + \eta_l ( k_{\bot} + \bar{k}_{\bot}) ]
\gamma_l (  k_{\bot} , \bar{k}_{\bot} )
 \nonumber
 \\
 &  & \hspace{-25mm} -
\int_{ \bar{k}_{\bot}^{\prime} } \int \frac{ d \bar{q} d \bar{\epsilon}}{(2 \pi )^2 } 
\delta ( | \bar{q} | -1 )  [ {\bf{F}}_l ( \bar{q} , i \bar{\epsilon} , 
 \bar{k}_{\bot}^{\prime} ) ]_{\alpha \alpha}
 \nonumber
 \\
 & &  \hspace{-23mm} \times
\frac{ \gamma_l (  k_{\bot} + \bar{k}_{\bot }^{\prime} ,   \bar{k}_{\bot} )
\gamma_l (  k_{\bot} , \bar{k}_{\bot}^{\prime} )  \gamma_l (  k_{\bot} +  \bar{k}_{\bot } + \bar{k}_{\bot}^{\prime} ,   - \bar{k}_{\bot}^{\prime} )  }{
 [ i  \bar{\epsilon}
 - \alpha  \bar{q}  - \tilde{\Delta}_l ( k_{\bot} , \bar{k}_{\bot}^{\prime} )]
 [ i  \bar{\epsilon}
 - \alpha  \bar{q}  - \tilde{\Delta}_l ( k_{\bot} + \bar{k}_{\bot} , \bar{k}_{\bot}^{\prime} )]
 } 
 \; ,
 \nonumber
 \\
 & &
\label{eq:gammaflow}
 \end{eqnarray}
with initial condition $\gamma_{l=0}(  k_{\bot} , \bar{k}_{\bot} )  =1$.
A graphical representation of Eq.~(\ref{eq:gammaflow}) is
shown in Fig.~\ref{fig:threevertex}.

{\it{Results.}}
Eqs.~(\ref{eq:flowr})-(\ref{eq:gammaflow}) form a closed system of flow 
equations for the rescaled self-energy at the FS 
$r_l ( k_{\bot})$, the flowing
anomalous dimension $\eta_l ( k_{\bot} )$, and the three-legged vertex
$\gamma_l ( k_{\bot} , \bar{k}_{\bot} )$.
Of course, these equations can only be solved numerically, but the
qualitative behavior of the solutions
can also be extracted analytically.
To begin with, let us establish the 
relation with  the perturbative Eq.~(\ref{eq:IE}).
We set $g_4=0$ from now on, because
the dominant renormalization of the FS is due to the
$g_2$-process.
In the simplest approximation, we set
$\gamma_l ( k_{\bot} , \bar{k}_{\bot} ) \approx 1$, 
$\eta_l ( k_{\bot} ) \approx 0$
and replace the flowing FS 
$\tilde{k}_{F,l} ( k_{\bot} ) = k_F ( k_{\bot}) / \Lambda - 
r_l ( k_{\bot} )$ by $k_F ( k_{\bot} ) / \Lambda $.
Then we obtain from Eqs.~(\ref{eq:r0flow}) and (\ref{eq:dotGamma})
to leading logarithmic order
\begin{equation}
 \frac{\delta k_F  ( k_{\bot} )}{\Lambda_0} 
 = [ ( 1 - g_2^2)^{-\frac{1}{2}} -1 ]  
 \int_{\bar{k}_{\bot}} 
 \tilde{\Delta} ( k_{\bot} , \bar{k}_{\bot} )
 \ln  |  \tilde{\Delta} ( k_{\bot} , \bar{k}_{\bot} ) |
 \; .
 \end{equation}
Expanding in harmonics we obtain as before
$t / t_0 = [ 1 + R ( t )]^{-1}$, but now with
 \begin{eqnarray}
 R ( t ) & \approx &    
 [ ( 1 - g_2^2)^{-\frac{1}{2}} -1 ]  
  \ln ( {\Lambda_0  }/{ | t|  } )
\; ,
 \label{eq:R2}
\end{eqnarray}
which reduces to Eq.~(\ref{eq:Rdef}) to leading order in $g_2$.  
Obviously, $R ( t ) \rightarrow \infty$ for $g_2 \rightarrow 1$,
indicating a confinement transition at strong coupling, where the renormalized
interchain hopping $t$ vanishes.
However, from our previous work
on two coupled chains \cite{Ledowski07}
we know that vertex corrections and
wave-function renormalizations
can possibly change this scenario.

\begin{figure}[tb]
 \vspace{5mm}
  \centering
\epsfig{file=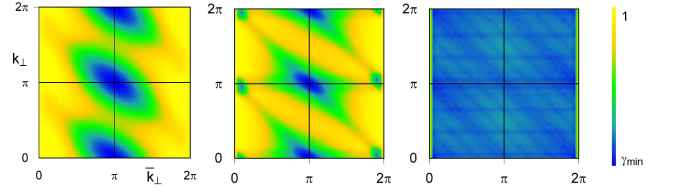,width=8.4cm}

  \caption{%
(Color online) Typical evolution of the vertex $\gamma_l ( k_{\bot} , \bar{k}_{\bot} )$
for different values of the flow parameter $l$. 
To evaluate the flow we have expanded in Eqs.~(\ref{eq:dotGamma}) and (\ref{eq:gammaflow}) up to $g_2^2$.
We have assumed a
harmonic bare FS with amplitude $t_0 / \Lambda_0 = 10^{-3}$
and bare coupling $g_2 =0.4$ 
From left to right 
$l= \frac{1}{2}l_c, l_c,3l_c$ and $\gamma_{\min}= 0.999989,0.981,0.85$,
where  the crossover scale $l_c$ (see text) can be approximated by
$l_c \approx - \ln (2 t_0 / \Lambda_0 )$ for small $g_2$.
}
  \label{fig:flowvertex}
\end{figure}

\begin{figure}[tb]
  \centering
\epsfig{file=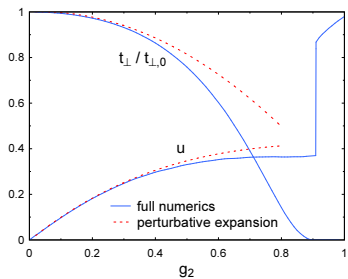,width=45mm}
  \vspace{-4mm}
  \caption{%
(Color online) 
Renormalized nearest neighbor interchain hopping 
$t_{\bot}$ and renormalized  
interaction $u = g_{\infty} ( \pi / a_{\bot} )$
as a function of the bare
interaction $g_2$ for  $\tilde{t}_0 = 2 t_{\bot,0} / (v_F \Lambda_0 ) = 0.1$ 
as obtained from the numerical solution
of Eqs.~(\ref{eq:dotGamma}) and (\ref{eq:gammaflow}).
For the perturbative curves we have expanded in these equations up 
to order $g_2^2$.}
  \label{fig:flowAu}
\end{figure}

\begin{figure}[tb]
  \centering
\epsfig{file=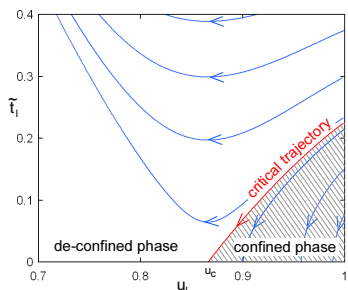,width=45mm}
  \vspace{-4mm}
  \caption{%
(Color online)
Projection of the RG flow  in the  $\tilde{t}$-$u$ plane.
The trajectories are obtained from the numerical solution of  Eq.(\ref{eq:uflowfinal}) and
$\tilde{t}_l = ( 1 - \bar{\eta}_l ) \tilde{t}_l$, where $\bar{\eta}_l$ is the
weighted FS average of the flowing anomalous dimension 
$\eta_l ( k_{\bot} )$  defined in the text.
}
  \label{fig:flowdiagram}
\end{figure}

To investigate this, let us first consider the
RG flow of the vertex $\gamma_l ( k_{\bot} , \bar{k}_{\bot} )$ numerically.
Representative results are shown in Fig.~\ref{fig:flowvertex}.
%
Obviously, 
the dependence of $\gamma_l ( k_{\bot} , \bar{k}_{\bot} )$
 on the fermionic momentum
$k_{\bot}$, which develops at intermediate scales $l$, is smoothed out again at
the scale $l_c$ where the
reduced cutoff $\Lambda_c = \Lambda_0 e^{-l_c} $ becomes comparable 
with the warping of the renormalized FS.
Defining the flowing dimensionless 
nearest neighbor interchain hopping $\tilde{t}_l$ via
$\tilde{k}_{F,l} = \bar{k}_F / \Lambda + \tilde{t}_l
 \cos ( k_{\bot} a_{\bot} ) + \ldots$, 
the crossover scale $l_c$ can be defined 
self-consistently via
$2 \tilde{t}_{l=l_c} = 1$.
%
%
%
Because of the weak  dependence
of  $\gamma_l ( k_{\bot} , \bar{k}_{\bot} )$
on  the fermionic momentum ${k}_{\bot}$, we may approximate
$\gamma_l ( k_{\bot} , \bar{k}_{\bot} ) \approx
\gamma_l ( 0 , \bar{k}_{\bot} ) \equiv \gamma_l (\bar{k}_{\bot} )  $.
Moreover, using the fact that close to the confinement transition $|\tilde{t}_l| \ll 1$,  
we obtain from
Eqs.~(\ref{eq:flowr})-(\ref{eq:gammaflow})
to leading order in $\tilde{t}_l$ 
the following RG flow equation for the effective interaction
 $g_l ( \bar{k}_{\bot}  ) = g_2 \gamma_l^2 ( \bar{k}_{\bot} )$,
\begin{equation}
 \partial_l g_l ( \bar{k}_{\bot} ) = -
 \frac{   4 \sin^2 ( \bar{k}_{\bot} a_{\bot}/2  )
   g_l ( \bar{k}_{\bot} ) u_l^2  \tilde{t}_l^2   }{
 \sqrt{ 1 - u_l^2} [ 1 + \sqrt{ 1 - u_l^2} ]^3}
 \; ,
 \label{eq:uflowfinal}
 \end{equation}
 with $u_l = g_l ( \pi / a_{\bot} )$.
Note that the flow of $g_l ( \bar{k}_{\bot}  )$ is driven by the
component  $ g_l ( \pi / a_{\bot} )$  of the interaction involving 
momentum transfer $\bar{k}_{\bot} = \pi / a_{\bot}$; 
in a simplified two-chain model 
this corresponds to the pair-tunneling process~\cite{Fabrizio93}.
The flow of the rescaled interchain hopping
 $\tilde{t}_l$ is determined by
 $\partial_l \tilde{t}_l =  ( 1 - \bar{\eta}_l ) \tilde{t}_l + O ( \tilde{t}_l^2 )$,
where  $\bar{\eta}_l$ is
the weighted FS average 
 $\bar{\eta}_l =
    2 
   \int_{ {k}_{\bot}}   
  \sin^2 (  {k}_{\bot}  a_{\bot} /2 )
 \eta_l( {k}_{\bot} )$ 
of the flowing anomalous dimension, which for small
$\tilde{t}_l$ can be approximated by
$ \eta_l( k_{\bot} ) =
  [  1 - g_l^2 ( k_{\bot} )  ]^{- \frac{1}{2}}  -1$.
From the numerical solution of these equations we find that,
for a given  $t_{\bot , 0} \ll v_F \Lambda_0$, there exists  a critical value of the
bare interaction $g_{2}$ where the renormalized $\tilde{t} = \tilde{t}_{\infty}$ 
vanishes, corresponding to a confinement 
transition to a state with vanishing interchain hopping $t_{\bot}$ and flat FS. 
In Fig.~\ref{fig:flowAu} we show the ratio $t_{\bot} / t_{\bot ,0}$
together with the renormalized interaction $u = g_{\infty} ( \pi / a_{\bot} )$ 
as a function of
the bare interaction for small $\tilde{t}_0 = 2 t_{\bot ,0} / (v_F \Lambda_0)$. 
A projection of the RG flow in the $\tilde{t}$-$u$ plane  
is shown in Fig.~\ref{fig:flowdiagram}.
At the confinement transition the system is certainly not a
Fermi liquid, because the anomalous dimension
$\eta_l ( k_{\bot} )$  is larger than unity and a
sharp FS cannot be defined~\cite{Kopietz01}.
The physical properties of the model in the confined phase
remain to be explored.

{\it{Summary and conclusions.}}
We have
shown that sufficiently strong interactions 
involving momentum transfers parallel to the chains can lead to a 
confinement transition in highly anisotropic
quasi-one-dimensional metals.
At the confinement transition the curvature
of the two disconnected sheets of the FS vanishes, so that
a coherent motion of the electrons perpendicular to the chains
is not possible and the electronic motion is one-dimensional.
Our calculation thus  supports the existence of a confined state
in quasi-one-dimensional metals, as originally suggested  by
Clarke, Strong, and Anderson \cite{Clarke94}.
In the confined state the system is a non-Fermi liquid with large anomalous dimension.
From the numerical solution of our  functional RG equations we found no
evidence for a truncation transition~\cite{Furukawa98,Ferraz03}, where
only certain sectors of the FS are washed out by interactions.

Because in this work we have  considered only spinless fermions 
and interactions
which do not transfer momentum between the two disconnected sheets of the FS,
our model (\ref{eq:Sdef}) is too simple  to
describe the competition between confinement and the
tendency to develop  some kind of 
long-range order, such as  charge-density or  spin-density waves.
In principle, spin fluctuations can be
taken into account with the help of another Hubbard-Stratonovich field,
but the analysis of the resulting  RG equations for the coupled 
boson-fermion model requires a substantial extension of our
calculation.

%

We thank A. Ferraz and F. H. L. Essler for useful discussions.

\end{document}